\documentclass[prd,preprintnumbers,nofootinbib]{revtex4} 
\usepackage{graphicx} 
\usepackage{amsmath}
\usepackage{amsfonts,amsbsy}
\usepackage{amssymb}

\def\gsim{ \,\, \vcenter{\hbox{$\buildrel{\displaystyle >}\over\sim$}}
 \,\,}

\def\be{\begin{equation}}
\def\ee{\end{equation}}
\def\bea{\begin{eqnarray}}
\def\eea{\end{eqnarray}}
\def\one{1\!\!1_R}
\def\Tr{{\rm tr}}

\begin{document}

\title{\bf Two-gluon correlations and initial conditions for small-$x$
  evolution}

\preprint{RBRC-898}

\author{Adrian Dumitru$^{a,b,c}$, Jamal Jalilian-Marian$^{b,c}$, Elena
Petreska$^{b,c}$}
\affiliation{
$^a$ RIKEN BNL Research Center, Brookhaven National
  Laboratory, Upton, NY 11973, USA\\
$^b$ Department of Natural Sciences, Baruch College, CUNY,
17 Lexington Avenue, New York, NY 10010, USA\\
$^c$ The Graduate School and University Center, City
  University of New York, 365 Fifth Avenue, New York, NY 10016, USA}

\begin{abstract}
We derive the effective action of the hard large-$x$ valence charges
up to fourth order in their density. Such non-Gaussian weight
functionals contribute at leading order in $N_c$ to the connected
two-gluon production diagrams which determine di-hadron correlations. The
corresponding diagrams are not necessarily (highly) suppressed by the
density of valence charges since their infrared divergences differ
from those obtained in a Gaussian theory.  Therefore, it appears
prudent to include such higher dimensional operators when determining
initial ensembles for JIMWLK evolution of higher $n$-point functions
of Wilson lines.
\end{abstract}

\maketitle

%---------------------------------------------------------------------------
\section{Introduction}

(Multi-) Particle production cross sections in hadronic collisions at
high energies are related to expectation values of traces of products
of Wilson lines which resum multiple scattering and high gluon density
effects. The evolution of such operator expectation values is
described by the JIMWLK functional renormalization group
equation~\cite{jimwlk}. In the large-$N_c$ limit, and in a Gaussian
approximation for the effective action, they reduce to the BK
equation~\cite{Kovchegov:1999yj} for the dipole forward scattering
amplitude.

The high-energy evolution equations require initial conditions at a
rapidity $Y=\log x_0/x =0$ ($x_0$ is often assumed to be about
$10^{-2}$). They have been derived by McLerran and
Venugopalan~\cite{MV} in the limit of an infinitely large nucleus. In
the MV model, the large-$x$ valence charges act as recoilless sources
for the soft, small-$x$ gluon fields. As $A\to\infty$ then, the
variance of the ``valence'' charge density $\sim A^{1/3}$ grows
large and the distribution of color charges should be Gaussian,
\be  \label{intro:MV}
S_{\rm MV} = \int d^2x_\perp \frac{{\rm tr}~\rho^2(x_\perp)}{\mu^2} =
\int d^2x_\perp \frac{\delta^{ab}\rho^a(x_\perp)\rho^b(x_\perp)}{2\mu^2}~~~~,~~~
\mu^2 \sim \frac{g^2 A}{\pi R^2}~.
\ee
It should be noted that high multiplicity proton-proton collisions may
also correspond to unusually high valence charge densities (see, for
example, ref.~\cite{Tribedy:2010ab}) and so would also be described
effectively by this action with $A_{\rm eff}\gg1$.

Nevertheless, in reality the mass number $A$ resp.\ the number of
valence charges is finite, in particular in (high-multiplicity) $pp$
as well as peripheral $AA$ collisions. It is therefore interesting to
consider extensions of the MV-model action involving higher powers of
the color charge density. In fact, Jeon and Venugopalan have
derived~\cite{JV} an ``odderon'' operator contribution $-d^{abc}
\rho^a \rho^b \rho^c/\kappa_3$, where the cubic coupling $\kappa_3\sim
g^3 (A/\pi R^2)^2$. Below, we shall show that at quartic order in
$\rho$ a contribution $\delta^{ab} \delta^{cd} \rho^a \rho^b \rho^c
\rho^d/\kappa_4$ to the effective action arises, with $\kappa_4 \sim
g^4 (A/\pi R^2)^3$.

Our motivation derives from the recent observation of a ``ridge'' in
two-particle correlations from high-multiplicity
$pp$~\cite{Khachatryan:2010gv} as well as heavy-ion
collisions~\cite{ridge_RHIC}. The ridge refers to a correlation which
extends over several units\footnote{It is presently not known whether
  the correlation simply satisfies boost invariance or whether it
  diminishes for rapidity intervals on the order of
  $1/\alpha_s$~\cite{Dumitru:2010iy,Kovchegov:1999ep}.} of
(relative) rapidity for particles with similar azimuthal angle,
$\Delta \phi \ll \pi$. For high-multiplicity $pp$ collisions at
$\sqrt{s}=7$~TeV the CMS collaboration found that the amplitude of the
``ridge'' correlation peaks about transverse momenta on the order of
2-3~GeV, a semi-hard scale. Ref.~\cite{Dumitru:2010iy} argued that
within the framework of high energy evolution this effect would
correspond to production of two small-$x$ gluons with relative
rapidity $\Delta y\gsim1$ from two evolution ladders which connect to
the same large-$x$ sources~\cite{CGCridge,DGLV2010}.

With a Gaussian action one finds that the two-particle distribution
minus the product of two single-particle distributions,
\be \label{intro:Correlation}
\frac{ \left\langle \frac{dN_2}{d^2pdy_p\,d^2qdy_q} \right\rangle}
{ \left\langle \frac{dN}{d^2pdy_p} \right\rangle
 \left\langle \frac{dN}{d^2qdy_q} \right\rangle} -1
~\sim~ \frac{1}{N_c^2-1}
\ee
is suppressed by a factor $1/(N_c^2-1)$ relative to the uncorrelated
part $\left\langle \frac{dN}{d^2pdy_p} \right\rangle \left\langle
\frac{dN}{d^2qdy_q} \right\rangle$. At this subleading order in $N_c$,
however, Gaussian factorization of the four-point function which
determines~(\ref{intro:Correlation}) is violated by JIMWLK
evolution~\cite{Dumitru:2010mv}.

This problem could be cured by a dynamically generated correlation
length (in the transverse plane) of small-$x$ gluon
fields~\cite{Kovner:2010xk}. For a discussion of how color charge
correlations develop as a hadron or nucleus is boosted to high
rapidity we refer to ref.~\cite{Mueller:2002pi}. Averaging of
$n$-point functions of Wilson lines with a local Gaussian would not
apply when those $n$ points are within one correlation length
of each other and consequently a connected contribution should arise
at leading order in $N_c$~\cite{Kovner:2010xk}. This can be seen also
from a model which assumes splitting of the four-point function into
two-point functions (``ladders'') with non-zero individual transverse
momenta~\cite{Levin:2011fb}.

Here, we show that a non-Gaussian initial distribution of valence
charge naturally arises when $A<\infty$ and that it leads to particle
correlations at leading order in $N_c$. Furthermore, that the infrared
behavior of the leading connected two-particle production diagram is
different from the case of a quadratic action. This fact may affect
the relative $A$-dependence of the correlations resulting from the
quadratic vs.\ the higher-order terms in the action.

We focus on the kinematic regime where the transverse momenta $p$, $q$
of the two produced gluons are somewhat larger than {\em but on the
  order of} the scale where evolution in rapidity is non-linear. For a
proton, this so-called saturation scale $Q_s(x)$ is expected to be on
the order of 1~GeV at $x\approx$ a few times $10^{-4}$, on average;
high-multiplicity collisions should correspond to configurations with
significantly higher parton densities. For nuclei, the valence charge
density should be boosted by a factor of $\sim A^{1/3}$. In this
kinematic regime, the transverse momenta of the produced gluons are,
to a significant part, due to the intrinsic transverse momentum from
the small-$x$ evolution ladders. Further, the relative azimuthal angle
of ${\bf p}$ and ${\bf q}$ is taken to be small, $\Delta\phi\ll\pi$.

When $\log p^2/Q_s^2 \gg 1$ and $\log q^2/Q_s^2 \gg 1$ the situation is
different. This case has been analyzed within the double-logarithmic
approximation in ref.~\cite{Bartels:2011qi}. They show that here a
single ``flip'' (or ``recombination'') between the two evolution
ladders is suppressed and that {\em two} such flips are required, one
below and the other above the rapidities of the produced {\em di}-jets. 
At very high energies such flips between ladders may occur at a hard
transverse momentum scale not too far from the hard dijet vertices
themselves~\cite{Bartels:2011qi}.

%---------------------------------------------------------------------
\section{Derivation of the effective action beyond quadratic order}
In this section we derive the form of the effective action for a
system of $k$ quarks in SU(3) following the methods developed
in~\cite{JV}. The probability to find this system in a representation
labeled by the integers $(m,n)$, also denoted as Dynkin labels, is
given by $d_{mn}\, N^{(k)}_{m,n}$ where 
\be e^{-S} \equiv d_{m,n}\,
N^{(k)}_{m,n} = d_{m,n} \bigg[G^{(k)}_{m,n} + G^{(k)}_{m+3,n} +
  G^{(k)}_{m,n+3} - G^{(k)}_{m+2,n-1} - G^{(k)}_{m-1,n+2} -
  G^{(k)}_{m+2,n+2} \bigg]
\label{eq:N_mul}
\ee
with       
\be
G_{k:\, m,n} = \frac{k!}
{\left(\frac{k+2m+n}{3}\right)!\,
\left(\frac{k-m+n}{3}\right)!\,
\left(\frac{k-m-2n}{3}\right)!
} ~.
\ee
We are interested in the representation with the largest weight and
so consider the limit $k\gg m,n\gg 1$. Thus, we expand the factorials
in each $G$ in powers of $1/k$
up to and including order $\sim {1\over k^3}$. The
leading terms of the form $\sim {m^j n^{l+1-j} \over k^l}$ can be
written as
\bea
S(m,n;k) \simeq {N_c\over k} C_2(m,n) - 
\frac{1}{3}\bigg({N_c\over k}\bigg)^2 C_3(m,n) +
\frac{1}{6}\bigg({N_c\over k}\bigg)^3 C_4 (m,n)
\eea
where $C_2$, $C_3$, $C_4$ are the
SU(3) Casimir operators for the representation $(m,n)$, given by (see
appendix)
\bea
C_2 (m,n) &\equiv & {1\over 3} \left( m^2 +n^2+mn\right)+
\left(m+n\right) \\
C_3 (m,n) &\equiv & {1\over 18} \left( m + 2 n +3\right) \left(n + 2 m
+3\right)\left(m-n\right) \\
C_4 (m,n) &\equiv & {1\over 9} \left(m^4 + n^4 + 2 m n^3 + 2 m^3
n + 3 m^2 n^2\right) + {2\over 3}
\left(m^3 + n^3 + 2 m^2 n + 2 m n^2\right) \nonumber \\
&&+ {1\over 6}\left(5 m^2 + 5 n^2 + 11 m n\right) 
- {1\over 2} \left(m + n\right) ~.
\eea
Before we rewrite the action $S(m,n;k)$ in terms of the color
charges $\rho$, it is perhaps useful to remind the reader of this
derivation and to illustrate it in the simpler case of SU(2)
spins~\cite{JV}.  In this case, the representation R is labeled by one
index, $l$, and there is only one independent Casimir defined as $C_2
(R) \one = \delta^{ab}\, L^a_R \, L^b_R = L^2_R$, where $L^a_R$ are
the generators of the representation $R$. For a large representation
we have $C_2 = \ell (\ell+1) \simeq \ell^2$ as the quadratic
Casimir. One can then write $\ell^2 = \ell_i\, \ell_i$ where $\ell_i$
is a vector describing spin and express the action in terms of
invariants formed by multiplying $\ell_i$'s (in this example, there is
only one invariant, $\ell^2$) as $S \simeq N_c \ell^2/k$. For large
$\ell$, one introduces a classical charge density per unit transverse
area, $\rho^i (x) =g\, \ell^i / \Delta^2 x$. Next, $k$ is expressed in
terms of number of valence quarks in a nucleus as $k = N_c\, A \,
\Delta^2 x / \pi R^2 $ which then leads to
\be S = {N_c\over k} \ell^2 \simeq \int d^2 x \, {\rho^a \rho^a
  \over 2 \mu^2} 
\ee
with $\mu^2 \equiv {g^2\, A\over 2 \pi R^2} $.

We start by defining the Casimirs for general $N_c$ in terms of the
generators $T_R^{a_i}$ of the representation $R$,
\be 
C_n (R) \one \equiv  F_R^{a_1,  \cdots , a_n} \,\, T_R^{a_1} \cdots
T_R^{a_n} ~,
\label{eq:C_def_matrix}
\ee
where $\one$ is the unit matrix in the representation $R$. Taking
the trace of both sides gives $C_n (R)$, the $n^{\rm th}$ Casimir of
the representation $R$:
\be 
C_n (R) = {1\over d_R} F_R^{a_1,  \cdots , a_n} \, \Tr\, T_R^{a_1}
\cdots T_R^{a_n} ~.
\label{eq:C_def_scalar}
\ee
The color tensors $F_R^{a_1, \cdots , a_n}$ are the most general
color invariant tensors one can construct out of the SU$(N_c)$
orthonormal basis tensors. For example, $F_R^{a_1,a_2} =
\delta^{a_1,a_2}$ for $n =2$, while for $n = 4$ these basis tensors
are given by $\delta^{ab}\, \delta^{cd}$ plus permutations,
$d^{abe}\,d^{cde}$ plus permutations, and $d^{abe}\, f^{cde}$ plus
permutations.  More explicitly,
\bea
C_2 (R) & = & {1\over d_R} \delta^{ab} \, \Tr\, T_R^a T_R^b  \\
C_3 (R) & = & {1\over d_R} d^{abc} \, \Tr\, T_R^a T_R^b T_R^c \nonumber \\
C_4 (R) & = & {1\over d_R} \big[\alpha (\delta^{ab} \delta^{cd} +
  \delta^{ac} \delta^{bd} + \delta^{ad} \delta^{bc}) +
\beta (d^{abe}d^{cde} +  d^{ace}d^{bde} + d^{ade}d^{bce})\big]    
\, \Tr\, T_R^a T_R^b T_R^c T_R^d \nonumber
\eea
Since the number of independent Casimirs of SU($N_c$) is equal to
its rank, $N_c -1$, there are only $2$ independent Casimirs for
SU(3). These are usually taken to be $C_2$ and $C_3$. This can be
seen explicitly by noting that for $N_c =3$ one has the constraint
\be
d^{abe}d^{cde} +  d^{ace}d^{bde} + d^{ade}d^{bce} = {1\over 3} 
\left(
\delta^{ab} \delta^{cd} +  \delta^{ac} \delta^{bd} + \delta^{ad}\delta^{bc}
\right)
\label{eq:constraint}
\ee
and $3\alpha + \beta = 1$ so that the quartic Casimir can be
written as the square of the quadratic invariant. To proceed, we
write the action in terms of $Q^a$, a $N_c^2 -1$ dimensional vector
related to the second Casimir via $|Q| = \sqrt{Q^a Q^a} \equiv
\sqrt{C_2}$; the vector $Q^a$ is analogous to the angular momentum
vector $\ell^a$ in our SU(2) example from above:
\bea
&& S \simeq \bigg[
{N_c\over k} Q^a Q^a - \bigg({N_c\over k}\bigg)^2 \,
\frac{d^{abc}}{3}\, Q^a Q^b Q^c  + 
\bigg({N_c\over k}\bigg)^3 \, {\delta^{ab} \delta^{cd} +  \delta^{ac}
  \delta^{bd} + \delta^{ad} \delta^{bc} \over 18} 
 Q^a Q^b Q^c Q^d\bigg]\nonumber 
\eea
We then follow McLerran-Venugopalan and define the color charge per unit area 
$\rho^a \equiv g\, Q^a / \Delta^2 x$ to finally arrive at
\bea
&&S[\rho(x) ] \simeq \int \, d^2 x \bigg[
{\delta^{ab} \,  \rho^a \rho^b \over 2 \mu^2} - {d^{abc}\, \rho^a \rho^b
  \rho^c \over \kappa_3 } +
{\delta^{ab} \delta^{cd} +  \delta^{ac} \delta^{bd} + \delta^{ad}
  \delta^{bc}  \over \kappa_4}
 \rho^a \rho^b \rho^c \rho^d\bigg]~.  \label{eq:quarticAction}
\eea
To write the action in this form we have used
eq.~(\ref{eq:constraint}) and $ k = N_c\, A {\Delta^2 x \over \pi
  R^2} $. This assumes that the large-$x$ sources correspond to
$N_c\,A$ valence quarks. On the other hand, if initial
conditions for small-$x$ evolution are set at, say, $x_0=0.01$ then the
number of ``valence'' charges would be bigger although still
parametrically proportional to $A$ and to $N_c$.

The couplings in this action are given by, parametrically,
\bea
\mu^2 &\equiv& {g^2 A \over 2 \pi R^2} \sim O (g^2 A^{1/3}) ~,\label{eq:mu2_gA}\\
\kappa_3 &\equiv& 3{g^3 A^2 \over (\pi R^2)^2} \sim O (g^3 A^{2/3})~, \\
\kappa_4 &\equiv& 18{g^4 A^3 \over (\pi R^2)^3} \sim O (g^4 A) ~.
\eea
The non-Gaussian terms in the action thus involve additional inverse powers
of $gA^{1/3}$. Also, the expressions from above are obtained by
averaging over the transverse (impact parameter) plane; at a fixed
distance $b$ from the center of a nucleus one should replace $A/\pi
R^2$ by the thickness function $T(b)$.

The action~(\ref{eq:quarticAction}) is coupled to the soft gauge
fields~\cite{jimwlk}. At leading order and in the classical
approximation this leads to the relation~(\ref{eq:Amu_rho}) below.

%---------------------------------------------------------------------------
\section{Two-point function and unintegrated gluon distribution}

\begin{figure}[htb]
\includegraphics[width=14mm]{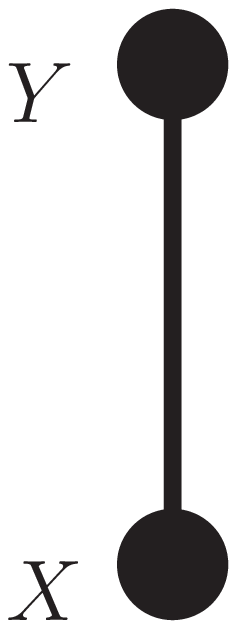}\hspace{1.5cm}
\includegraphics[width=30mm]{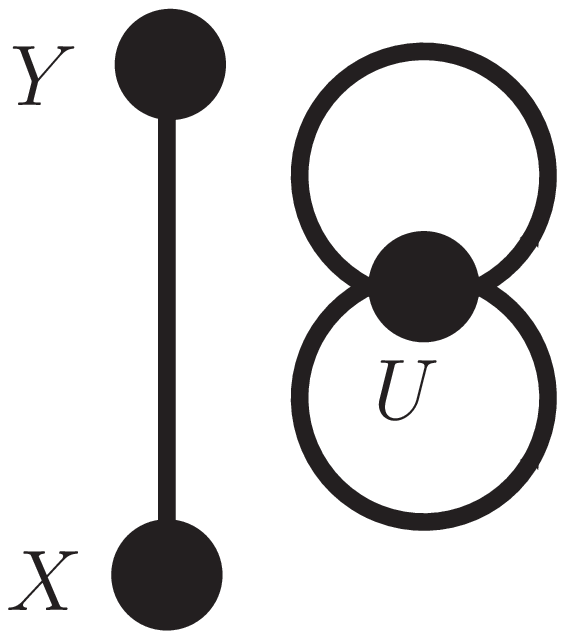}\hspace{1.5cm}
\includegraphics[width=24mm]{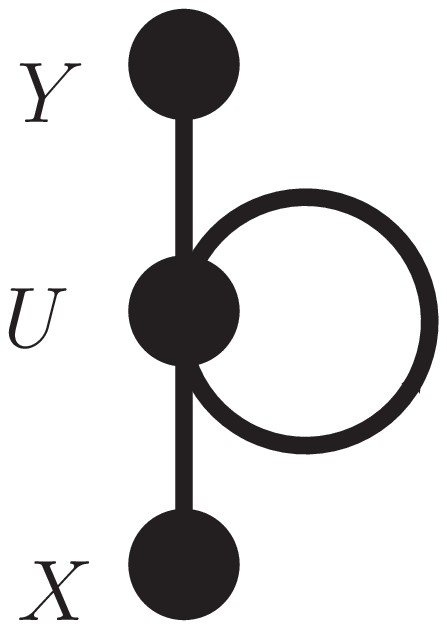}
\caption{Two-point function at leading order in $1/\kappa_4$.}
\label{fig:2pt_Pert}
\end{figure}
In order to compute the contribution of the quartic term to a physical
observable, such as single-inclusive hadron production in the forward
region~\cite{adjjm-prl}, one needs to perform the color averaging of
the dipole operator with this new action: within the Glauber-Mueller
approach the expectation value of the full dipole operator $\langle\Tr
\; V_x V^\dagger_y\rangle$, where $V$ denotes a Wilson line, resums
multiple scattering effects on a dense
target~\cite{Kovchegov:1999yj}. This can in principle be done
numerically using lattice gauge theory
methods~\cite{lattJIMWLK}. Nevertheless, it is useful to consider a
limit where this expectation value can be evaluated analytically, such
as the dilute limit of the dipole cross section given by the two-point
function of color charges. Even then, one can not perform the
integration over $\rho$ analytically when the action is not
quadratic. We therefore resort to a perturbative expansion in
$1/\kappa_4$ and keep only the first term in the expansion of the
exponential of the quartic term. We then compute the two-point
function of the color charge density of hard sources which is related
to the unintegrated gluon distribution. To first non-trivial order in
$1/\kappa_4$,
\be \label{eq:VEV_pert_kappa4}
\langle O[\rho] \rangle \equiv \frac{\int {\cal D}\rho \; O[\rho]\;
e^{-S_G[\rho]} 
\left[1 - \frac{1}{\kappa_4}\int d^2u\;
  \rho^a_u \rho^a_u \rho^b_u \rho^b_u\right]}
{\int {\cal D}\rho \;e^{-S_G[\rho]}
\left[1- \frac{1}{\kappa_4}\int d^2u\;
  \rho^a_u \rho^a_u \rho^b_u \rho^b_u\right]\;
}~.
\ee
Here, $S_G$ denotes the Gaussian action. For the two-point
function $O= \rho^a_x
\rho^b_y$, the possible contractions are shown diagrammatically in
fig.~\ref{fig:2pt_Pert}; for the denominator of
eq.~(\ref{eq:VEV_pert_kappa4}) one amputates the points $x$ and
$y$. We compute the functional integral in lattice regularization,
i.e.\ we approximate the two-dimensional transverse space by a lattice
with $N_s$ sites of area $\Delta^2 x$. The two-point function
becomes\footnote{The right-hand-side of this expression is to be
  understood in lattice notation: $x$ and $y$ are discrete points and
  $\delta_{xy}$ is a Kronecker symbol.}
\bea  \label{eq:q_2pt}
\langle \rho^a_x \rho^b_y \rangle &=& \mu^2 
\frac{\delta^{ab} \delta_{xy}}{\Delta^2 x}\,
\frac{1 - \frac{\mu^4}{\kappa_4} (N_c^2+1) (N_c^2-1) \frac{N_s}{\Delta^2 x} 
- 4 \frac{\mu^4}{\kappa_4} (N_c^2+1) \frac{1}{\Delta^2 x}}
{1 - \frac{\mu^4}{\kappa_4} (N_c^2+1) (N_c^2-1) \frac{N_s}{\Delta^2 x}}
~.
\eea
The disconnected contribution exhibits both a UV ($\Delta^2 x \to
0$) as well as a IR ($N_s\to \infty$) divergence but appears both in
the numerator and in the denominator and so cancels as usual.  The
third term in the numerator is due to the tadpole diagram from
fig.~\ref{fig:2pt_Pert} which renormalizes the ``bare'' parameter $\mu^2$:
\be \label{eq:mu_ren}
\tilde\mu^2 \equiv \mu^2 \left(
1 - 4 \frac{\mu^4}{\kappa_4} \frac{N_c^2+1}{\Delta^2 x} \right)~.
\ee
This renormalization of $\mu^2$ will be applied to all diagrams to
absorb the insertion of a tadpole into any line.

Thus, the two-point function now reads (in continuum notation)
\be  \label{eq:2ptRhoCont}
\langle \rho^a_x \rho^b_y \rangle = \tilde\mu^2 \delta^{ab}
\delta(x-y) ~~
\leftrightarrow~~
\langle \rho^{*\,a}(k) \rho^b(k') \rangle = \tilde\mu^2 \delta^{ab}
\, (2\pi)^2\delta(k-k') ~.
\ee
To relate~(\ref{eq:2ptRhoCont}) to the
unintegrated gluon distribution we note that at leading order and in
covariant gauge the field generated by a fast particle moving in the
positive $z$-direction is related to the charge density by
\be \label{eq:Amu_rho}
A^\mu(x) \equiv \delta^{\mu +} \alpha(x) =
     g \, \delta^{\mu +} \delta(x^-) \frac{1}{\nabla^2_\perp} \rho(x)~,
\ee
where again $x$ denotes a transverse coordinate. This field also
satisfies $A^-=0$ and so the only non-vanishing field-strength is
$F^{+i} = - \partial^i \alpha$.  In momentum space we have the
relation $k^2\alpha(k)=g\rho(k)$.  Thus, we define the unintegrated
gluon distribution $\Phi(k^2)\sim k^2 \left< F^{+i}F^{+i}\right>$ via
\be
\left< {\rho^*}^a(k) \rho^b(k')\right> = \frac{1}{\alpha_s} 
   \frac{\delta^{ab}}{N_c^2-1} (2\pi)^3 \delta(k-k') \, \Phi(k^2)~.
\ee
With this convention, $\Phi(k^2) = \alpha_s (N_c^2-1) \tilde\mu^2/(2\pi)$.

%---------------------------------------------------------------------------
\section{Four-point function}

\begin{figure}[htb]
\includegraphics[width=21mm]{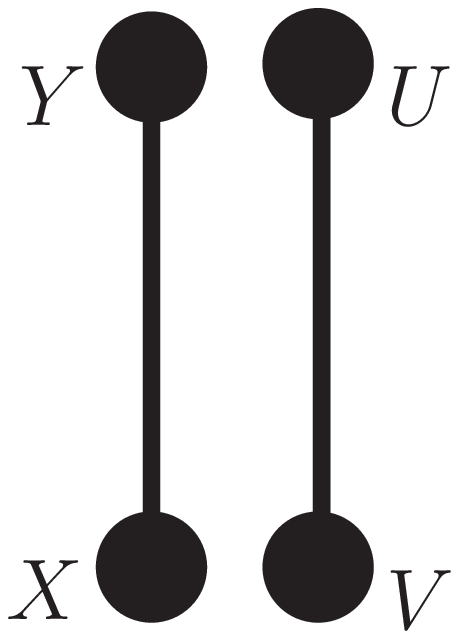}\hspace{1cm}
\includegraphics[width=30mm]{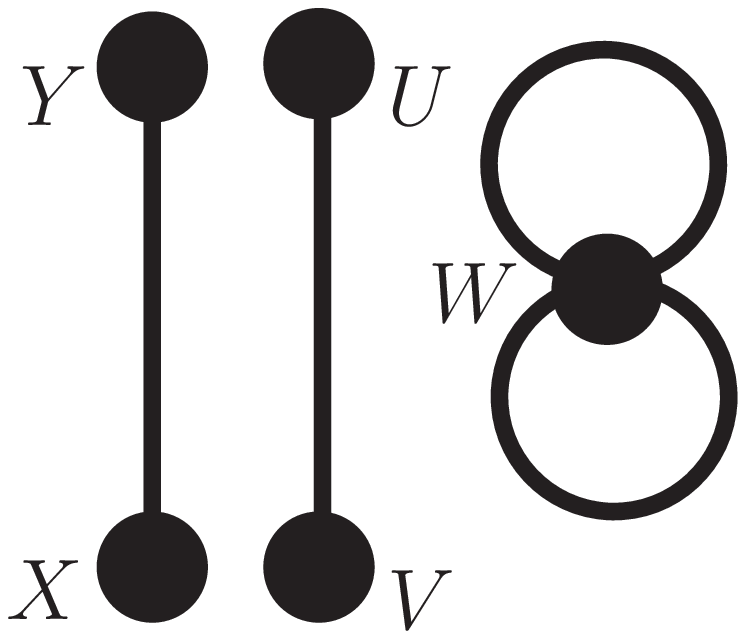}\hspace{1cm}
\includegraphics[width=24mm]{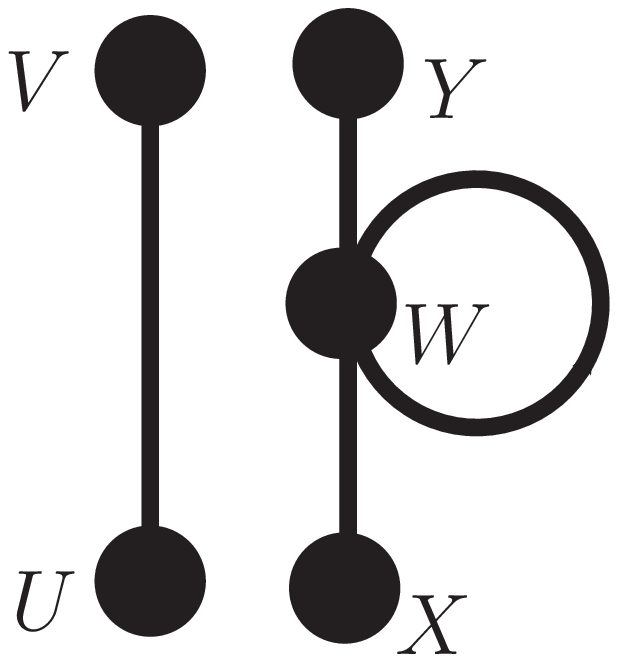}\hspace{1cm}
\includegraphics[width=24mm]{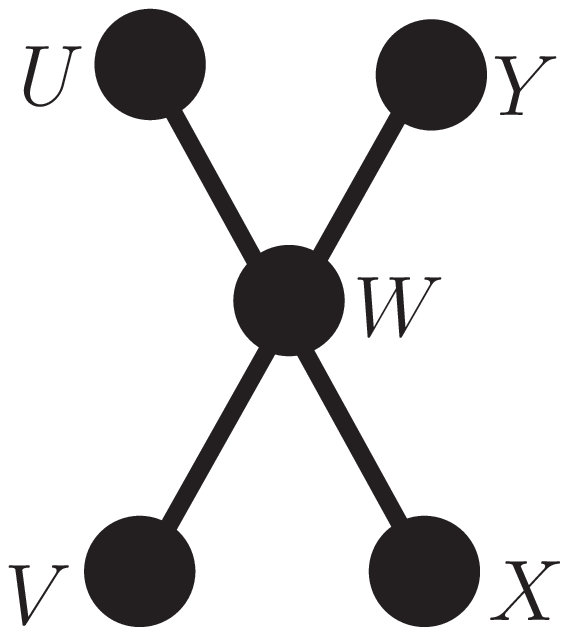}
\caption{Four-point function at leading order in $1/\kappa_4$.}
\label{fig:4pt_Pert}
\end{figure}
The four-point function to first non-trivial order in $1/\kappa_4$ is
given by
\be \label{eq:4ptVEV_pert_kappa4}
\langle \rho^a_x\rho^b_y\rho^c_u\rho^d_v \rangle \equiv 
\frac{\int {\cal D}\rho \; e^{-S_G[\rho]} \;
\rho^a_x\rho^b_y\rho^c_u\rho^d_v\;
\left[1 - \frac{1}{\kappa_4}\int d^2w\;
  \rho^a_w \rho^a_w \rho^b_w \rho^b_w\right]}
{\int {\cal D}\rho \;e^{-S_G[\rho]}
\left[1- \frac{1}{\kappa_4}\int d^2w\;
  \rho^a_w \rho^a_w \rho^b_w \rho^b_w\right]\;
}~.
\ee
The different type of contractions which arise at this order are shown
in Fig.~\ref{fig:4pt_Pert}. Eq.~(\ref{eq:4ptVEV_pert_kappa4}) then
gives
\bea
\langle \rho^a_x\rho^b_y\rho^c_u\rho^d_v \rangle &=&
\frac{1}{1 - \frac{\mu^4}{\kappa_4} (N_c^2+1) (N_c^2-1) \frac{N_s}{\Delta^2 x}}
\nonumber\\
& & 
\left\{
\frac{\mu^4}{(\Delta^2 x)^2} \left(
  \delta^{ab}\delta_{xy}\delta^{cd}\delta_{uv} +
  \delta^{ac}\delta_{xu}\delta^{bd}\delta_{yv} +
  \delta^{ad}\delta_{xv}\delta^{bc}\delta_{yu}
\right) \left[
1 - \frac{\mu^4}{\kappa_4} (N_c^2+1) (N_c^2-1) \frac{N_s}{\Delta^2 x} 
- 8 \frac{\mu^4}{\kappa_4} (N_c^2+1) \frac{1}{\Delta^2 x}
\right]
\right. \nonumber\\
& & \hspace{.5cm}
\left. - 8 \frac{\mu^8}{\kappa_4} \frac{1}{(\Delta^2 x)^3} 
\left(
  \delta^{ab}\delta^{cd} +
  \delta^{ac}\delta^{bd} +
  \delta^{ad}\delta^{bc}
\right) \delta_{xy} \delta_{xu} \delta_{uv}
\right\} ~.  \label{eq:4pt_lat}
\eea
\begin{figure}[htb]
\includegraphics[width=34mm]{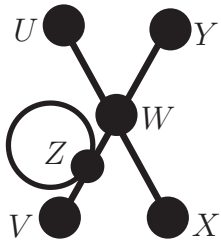}
\caption{Diagram at order $1/\kappa_4^2$ which renormalizes $\mu^2$ in
  the connected contribution to the four-point function.}
\label{fig:4pt_Pert_ren}
\end{figure}
The second line in this expression corresponds to the sum of the first
three diagrams shown in Fig.~\ref{fig:4pt_Pert}: the second term due
to the disconnected diagram again cancels against the overall
normalization factor once the latter is expanded to leading order in
$1/\kappa_4$. Also, the third term from the second line is again
absorbed into the renormalized value of $\mu^2$ as
in~(\ref{eq:mu_ren}). In fact such a factor of
$1-16\frac{\mu^4}{\kappa_4} (N_c^2+1) \frac{1}{\Delta^2 x}$ appears
at order $1/\kappa_4^2$ from diagrams of the type shown in
Fig.~\ref{fig:4pt_Pert_ren} which induce the shift $\mu^8 \to
\tilde\mu^8$ in the last line of eq.~(\ref{eq:4pt_lat}). Thus, the
four-point function to order $1/\kappa_4$ finally becomes (switching
again from lattice to continuum notation)
\bea
\langle \rho^a_x\rho^b_y\rho^c_u\rho^d_v \rangle &=&
\tilde\mu^4 \left[
  \delta^{ab}\delta^{cd}\delta(x-y)\delta(u-v)
    \left(1-8\frac{\tilde\mu^4}{\kappa_4}\delta(x-u)\right) +
  \delta^{ac}\delta^{bd}\delta(x-u)\delta(y-v) 
    \left(1-8\frac{\tilde\mu^4}{\kappa_4}\delta(x-y)\right)
    \right. \nonumber\\
& & \left. ~~ +
  \delta^{ad}\delta^{bc}\delta(x-v)\delta(y-u)
    \left(1-8\frac{\tilde\mu^4}{\kappa_4}\delta(x-y)\right) 
\right]  \label{eq:4pt_cont}
\eea
In momentum space,
\bea
\langle \rho^{*\,a}(k_1) \rho^{*\,b}(k_2) \rho^c(k_3) \rho^d(k_4) \rangle &=&
 \nonumber\\
& & \hspace{-2cm}
(2\pi)^4\, \tilde\mu^4 \bigg[
  \delta^{ab}\delta^{cd}\delta(k_1+k_2)\delta(k_3+k_4) +
  \delta^{ac}\delta^{bd}\delta(k_1-k_3)\delta(k_2-k_4) +
  \delta^{ad}\delta^{bc}\delta(k_1-k_4)\delta(k_2-k_3)
     \nonumber\\
& & \hspace{-2cm} \left. ~~~~~~~~~~ -
    \frac{2}{\pi^2} \frac{\tilde\mu^4}{\kappa_4}
    \left(\delta^{ab}\delta^{cd} + \delta^{ac}\delta^{bd} + \delta^{ad}\delta^{bc}
    \right)
    \delta(k_1+k_2-k_3-k_4)
\right]~.   \label{eq:4ptMom_cont}
\eea
%

%---------------------------------------------------------------------
\section{Correlated two-gluon production}

\begin{figure}[htb]
\includegraphics[width=12cm]{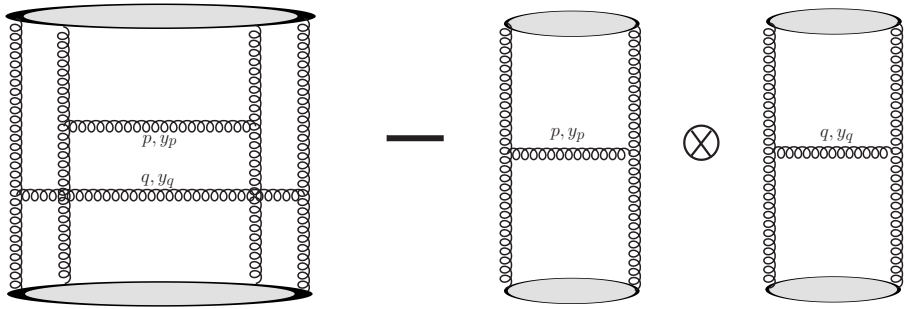}
\caption{Correlated production of two gluons with small relative
  azimuth, $\Delta\phi\ll\pi$.}
\label{fig:2gCorr}
\end{figure}
We now turn to the color factor for correlated production of two
gluons as illustrated in Fig.~\ref{fig:2gCorr}. We focus, in
particular, on the so-called ``ridge'' kinematics where the gluons are
separated by a rapidity interval $\Delta\eta\gsim1$ but where their
transverse momenta are nearly parallel so that their relative azimuth
$\Delta\phi\ll\pi$. We assume that in this kinematic regime the
dominant contribution arises from a process where the small-$x$ gluons
are produced from two separate ladders which connect to the same hard
sources, though~\cite{CGCridge}.

It was pointed out in ref.~\cite{Dumitru:2010mv} that the cross
section for this process involves a product of two four-point
functions (projectile and target, respectively) such as the one
discussed above. Further, that for a Gaussian action the correlated
contribution which is left over after one subtracts the square of the
single inclusive cross section is suppressed by $\sim1/N_c^2$.

The diagram from Fig.~\ref{fig:2gCorr} corresponds to~\cite{Dumitru:2010mv}
\be \label{eq:dN2_part}
f_{e aa^\prime} f_{e^\prime b b^\prime}
f_{ecc^\prime} f_{e^\prime dd^\prime} 
\Big< \rho^{*\,a} (k_2) \rho^{*\, b}(k_4)
\rho^c(k_1) \rho^d (k_3)\Big> 
\left< \rho^{*\,a^\prime}(p-k_2)
\rho^{*\, b^\prime}(q-k_4)
\rho^{c^\prime}(p-k_1) \rho^{d^\prime}(q-k_3)\right>
\ee
with $e$ and $e'$, respectively, the color indices of the two produced
gluons; the ladder momenta $k_i$ will be integrated over.
The (squared) single-inclusive cross section is obtained from
\bea
& &
f_{e aa^\prime} f_{e^\prime b b^\prime}
f_{ecc^\prime} f_{e^\prime dd^\prime} 
\Big< \rho^{*\,a}(k_2) \rho^c(k_1)\Big> 
\Big< \rho^{*\,b}(k_4) \rho^d(k_3)\Big> 
\left<\rho^{*\,a^\prime}(p-k_2) \rho^{c^\prime}(p-k_1) \right>
\left<\rho^{*\,b^\prime}(q-k_4) \rho^{d^\prime}(q-k_3)\right>  \nonumber\\
&=& (2\pi)^4\, N_c^2 (N_c^2-1)^2 \; (\pi R^2)^2\; \tilde\mu^8 \; 
\delta(k_1-k_2)\, \delta(k_3-k_4)~.  \label{eq:ColorSingle}
\eea
Translational invariance in the transverse plane brings along a factor
of $\delta(k=0)=\pi R^2/(2\pi)^2$ for each of the two produced gluons.
Using~(\ref{eq:mu2_gA}) the prefactor is of order 
\be \label{eq:SI2scaling}
\mathrm{(single~inclusive)}^2 \sim N_c^2 (N_c^2-1)^2
\;(\pi R^2)^2\; \left(\frac{g^2 A}{\pi R^2}\right)^4 
\sim \left(g A^{1/3}\right)^8 N_c^2 (N_c^2-1)^2 ~.
\ee

\begin{figure}[htb]
\begin{center}
\includegraphics[width=0.2\textwidth]{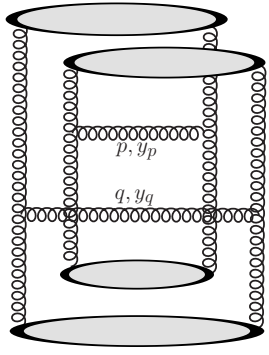}
\end{center}
\caption{\label{fig:diagPLB} Two-gluon production diagram
  which gives an angular collimation about $\Delta\phi=0$ over several
  units in relative rapidity, $|y_p-y_q|\gsim1$ (the so-called
  ``ridge'')~\cite{Dumitru:2010iy}. Such diagrams arise from a
  Gaussian action and only involve the two-point function
  (unintegrated gluon distribution); they are suppressed by
  $\sim1/(N_c^2-1)$ relative to uncorrelated production.}
\end{figure}
The remaining contractions in eq.~(\ref{eq:dN2_part}) generate
connected diagrams corresponding to correlated production. In the
limit $\kappa_4\to\infty$ they are suppressed by a factor
$1/(N_c^2-1)$. For example, for the diagram from fig.~\ref{fig:diagPLB},
\bea
& &
f_{e aa^\prime} f_{e^\prime b b^\prime}
f_{ecc^\prime} f_{e^\prime dd^\prime} 
\Big< \rho^{*\,a}(k_2) \rho^d(k_3)\Big> 
\Big< \rho^{*\,b}(k_4) \rho^c(k_1)\Big> 
\left<\rho^{*\,a^\prime}(p-k_2) \rho^{c^\prime}(p-k_1) \right>
\left<\rho^{*\,b^\prime}(q-k_4) \rho^{d^\prime}(q-k_3)\right>  \nonumber\\
&=& (2\pi)^6\, N_c^2 (N_c^2-1) \; \pi R^2\; \tilde\mu^8 \; 
\delta(k_1-k_2)\, \delta(k_3-k_4)\, \delta(k_1-k_4)~.  \label{eq:PLBdiag}
\eea
Due to the additional $\delta$-function as compared to
eq.~(\ref{eq:ColorSingle}) there is, of course, only one
single independent ladder momentum to integrate over. This diagram
will therefore depend on the relative azimuthal angle between
${\bf p}$ and ${\bf q}$; it is proportional to
\be \label{eq:prefactor_Gaussian}
\mathrm{Fig.\,5}~ \sim ~
N_c^2 (N_c^2-1) \; \pi R^2\; \frac{\tilde\mu^8}{p^2 q^2}
\int \frac{dk^2}{k^4} \frac{1}{(p-k)^2} \frac{1}{(q-k)^2} ~.
\ee
Its contribution is therefore maximal at $\Delta\phi=0$ and minimal at
$\Delta\phi=\pi/2$~\cite{Dumitru:2010iy}. Also, the integral is
quadratically divergent from the region $k^2\ll p^2, q^2$ and ${\bf
  k}\sim{\bf p}\sim{\bf q}$. If a fixed non-perturbative cutoff
$\Lambda^2$ is used then this contribution is $\sim A^2$, i.e.,
suppressed by $\sim A^{-2/3}$ as compared to uncorrelated production,
eq.~(\ref{eq:SI2scaling}). On the other hand, if such cutoff is in
fact provided by the non-linear saturation scale~\cite{CGCridge},
$Q_s^2 \sim A^{1/3}$, then~(\ref{eq:PLBdiag}) is of order $\sim
A^{5/3}$, down by $A^{-1}$ as compared to uncorrelated
production\footnote{Note that in practice this $\sim1/A$ suppression
  of the correlated contribution is usually removed by multiplying
  with the multiplicity per unit rapidity, $dN/dy$.}.

\begin{figure}[htb]
\begin{center}
\includegraphics[width=0.2\textwidth]{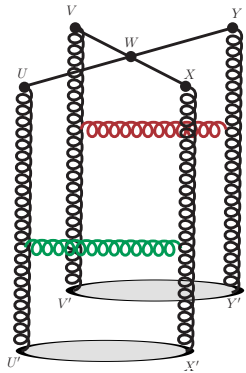}
\end{center}
\caption{\label{fig:CorrQuarticAction} Diagram for correlated
  two-gluon production from the quartic action. This contribution is
  of the same order in $g^2$ and $A$ as that from
  fig.~\ref{fig:diagPLB} but enhanced by a factor of $N_c^2-1$.}
\end{figure}
For finite $\kappa_4$, however, we can see from
eq.~(\ref{eq:4pt_cont},\ref{eq:4ptMom_cont}) that the term
$\sim \delta^{ac} \delta^{bd}$ produces the same overall color
factor as the square of the single-inclusive cross section. The leading
order (in $1/\kappa_4$) connected contribution of order $N_c^2 (N_c^2-1)^2$ 
to~(\ref{eq:dN2_part}) is shown diagrammatically in
fig.~\ref{fig:CorrQuarticAction} and is given by\footnote{The overall
  factor of 2 accounts for projectile $\leftrightarrow$ target flip.}
\bea
& &
-2 (2\pi)^8 \frac{2}{\pi^2} \frac{\tilde\mu^8}{\kappa_4} \tilde\mu^4 \;
\delta^{ac}\delta^{bd}\delta^{a'c'}\delta^{b'd'}
f_{e aa^\prime} f_{e^\prime b b^\prime}
f_{ecc^\prime} f_{e^\prime dd^\prime} \;
\delta(k_1-k_2) \; \delta(k_3-k_4) \; \delta(k_1+k_3-k_2-k_4) \nonumber\\
&=& 
-(2\pi)^6\, \frac{4}{\pi^2} \; N_c^2 (N_c^2-1)^2 \; \pi R^2\;
\frac{\tilde\mu^{12}}{\kappa_4} \;
\delta(k_1-k_2)\; \delta(k_3-k_4)~.  \label{eq:TwoPart_kappa4}
\eea
The contribution from this diagram to the two-particle spectrum is
\bea
&\sim& - N_c^2 (N_c^2-1)^2 \; \pi R^2\;
\frac{\tilde\mu^{12}}{\kappa_4} \;
\frac{1}{p^2\, q^2}
\int \frac{dk^2}{k^2} \frac{1}{(p-k)^2} 
\int \frac{dk^{\prime\,2}}{k^{\prime\,2}} \frac{1}{(q-k')^2} \nonumber\\
&\sim &
- N_c^2 (N_c^2-1)^2 \; \pi R^2\;
\frac{\tilde\mu^{12}}{\kappa_4} \;
\left(\frac{1}{p^2\, q^2}\right)^2
\log\frac{p^2}{\Lambda^2}\;  \log\frac{q^2}{\Lambda^2}~.
\eea
Thus, this contribution is independent of the azimuthal angle
$\Delta\phi$ between ${\bf p}$ and ${\bf q}$. Also, its sensitivity to
the infrared cutoff is only logarithmic and would therefore not
modify the $A$-dependence of the prefactor.

From~(\ref{eq:mu2_gA}) then the prefactor
of~(\ref{eq:TwoPart_kappa4}) is of order
\be \label{eq:prefactor_kappa4}
\mathrm{Fig.\,6}~ \sim ~ N_c^2 (N_c^2-1)^2 \; \pi R^2\; 
\left( \frac{g^2 A}{\pi R^2} \right)^6
\frac{1}{g^4} \left( \frac{\pi R^2}{A}\right)^3 
\sim
g^8 A^{5/3} \; N_c^2 (N_c^2-1)^2 ~.
\ee
This is of the same order in $g^2$ and $N_c$ as the uncorrelated
contribution~(\ref{eq:SI2scaling}) but suppressed by one power of the
mass number $A$. However, it is more relevant to compare the
contribution from the quartic interaction to the correlation,
eq.~(\ref{eq:prefactor_kappa4}), to that from the Gaussian action,
eq.~(\ref{eq:prefactor_Gaussian}). If the latter is cut off at a
constant, $A$-independent scale $\Lambda^2$ then it is parametrically
enhanced by a factor of $A^{1/3}$ and the corrections from the quartic
term in the action can be computed perturbatively (near $A=\infty$) by
an expansion in $1/\kappa_4$ (as we have done). One may expect
that such corrections are large when $A^{1/3}\simeq6$ since they are
enhanced by a factor of $N_c^2-1$.

On the other hand, if the infrared cutoff $\sim Q_s^2\sim
A^{1/3}$ then the non-Gaussian action generates ``corrections'' to
two-particle correlations which appear at the same order in $A$ as the
contribution from the Gaussian action (up to logarithms of $A$); in
addition, those ``corrections'' are enhanced by a large color factor
$N_c^2-1$. This case requires a non-perturbative calculation of the
correlation to all orders in $1/\kappa_4$.

\begin{figure}[htb]
\begin{center}
\includegraphics[width=0.25\textwidth]{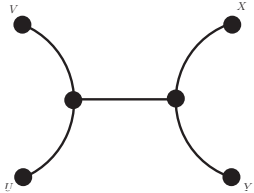}
\end{center}
\caption{\label{fig:4ptKappa3} Connected contribution to the
  four-point function at order $1/\kappa_3^2$.}
\end{figure}
We conclude this section by noting that the ``odderon'' operator can
also contribute to the connected two-particle production cross section
at order $\sim 1/\kappa_3^2$. This contribution is obtained by
expanding the exponential of the odderon operator to second
order. Performing the averaging of the four-point function shown in
fig.~\ref{fig:4ptKappa3} over the hard color sources like in
eq.~(\ref{eq:4ptVEV_pert_kappa4}) results in five Wick
contractions and is of order $\mu^{10}/\kappa_3^2$. Multiplying by the
leading $\sim\kappa_3^0$ contribution from the target side gives, in
all:
\be
+ \pi R^2\; N_c^2 (N_c^2-1) \left(N_c-\frac{4}{N_c} \right) \;
\frac{\tilde\mu^{14}}{\kappa_3^2} ~ \sim  ~
+ g^8 A^{5/3}\; N_c^2 (N_c^2-1) \left(N_c-\frac{4}{N_c} \right)~.
\ee
This is of the same order in $A$ as the contribution $\sim
1/\kappa_4$ from eq.~(\ref{eq:prefactor_kappa4}) although it is
suppressed by one power of $N_c$. It appears sensible that numerical
solutions of JIMWLK evolution also include the odderon operator in the
action for the initial condition. On the other hand, terms of order
$\sim\rho^6$ do not matter for the four-point function (they do matter
for the six-point function and higher), they simply renormalize $\kappa_4$.

%---------------------------------------------------------------------
\section{Summary} \label{sec:Summary}
In summary, we have computed corrections to the quadratic MV-model
action up to quartic order in the density of hard (large-$x$)
sources. Such terms are accompanied by additional powers of
$1/gA^{1/3}$. This action could be employed to determine initial
conditions for the JIMWLK high-energy evolution of various $n$-point
operator expectation values. We have motivated this by showing that at
leading order in the quartic coupling this action gives a connected
contribution to two-gluon production (i.e., to correlations) already
at leading order in $N_c$, enhanced by a factor of $N_c^2-1$ over the
connected contribution from a (local) quadratic action. Such non-Gaussian
actions may also be important for applications to dijet
production with transverse momenta not very far above the saturation
scale~\cite{Dumitru:2010ak}.

%---------------------------------------------------------------------
\begin{acknowledgments}
We thank S.\ Jeon for extensive discussions on the derivation
of the MV action in ref.~\cite{JV} and A.~Kovner for communications
which initiated this work.
We gratefully acknowledge support by the DOE Office of Nuclear Physics
through Grant No.\ DE-FG02-09ER41620, from the ``Lab Directed Research
and Development'' grant LDRD~10-043 (Brookhaven National Laboratory),
and from The City University of New York through the PSC-CUNY Research
Award Program, grants 63382-0041 (A.D.) and 63404-0041 (J.J-M.).
\end{acknowledgments}

%---------------------------------------------------------------------
\section{Appendix} \label{sec:CasimirAppendix}

In this appendix we illustrate how one can express the values of
various Casimirs in terms of the Dynkin labels $n$ and $m$ of SU(3).
The representation $(m,n)$ is reached by multiplying the $(1,0)$
fundamental representation $k$ times. The corresponding Young tableau
thus has $k$ boxes in $N=3$ rows. The first row has
$h_1=m+n+(k-m-2n)/3$ boxes; the second row has $h_2=n+(k-m-2n)/3$
boxes; and the last row has $h_3=(k-m-2n)/3$ boxes.

It is useful to define the set $n_i'$ for the U(N) group from these
``row lengths'' $h_i$ via $n_i'=h_i+(N+1)/2-i$:
\bea
n_1' &=& m+n+\frac{k-m-2n}{3} +2-1 = \frac{k+2m+n}{3} + 1 \\
n_2' &=& n+\frac{k-m-2n}{3} +2-2 = \frac{k-m+n}{3} \\
n_3' &=& \frac{k-m-2n}{3} +2-3 = \frac{k-m-2n}{3} - 1 \\
\sum_i n_i' &=& k ~.
\eea
To pass from U(3) to SU(3) one needs to shift $n_i=n_i'-\sum_j n_j'/N$ to get 
\bea
n_1 &=& \frac{2m+n}{3} + 1 \\
n_2 &=& \frac{-m+n}{3} \\
n_3 &=& \frac{-m-2n}{3} - 1 ~.
\eea
Therefore,
\bea
\sum_i n_i^2 &=& = \frac{2}{3} \left( m^2 +n^2+mn\right) + 2 (m+n+1) \\
\sum_i n_i^4 &=& \frac{2}{9} \left(3+m^2+n^2+3n+3m+mn\right)^2   ~.
\eea
The quadratic casimir is constructed from $n_i$
via~\cite{Dubath:2002qv}\footnote{Note that
their normalization of the generators is different from ours so that
their expressions for the Casimirs need to be divided by 2.}
\be  \label{eq:C2_SU3_Dubath}
C_2 (m,n) = \sum_i n_i^2 - \frac{N(N^2-1)}{12} = \frac{1}{3} \left( 
m^2 +n^2+mn\right)+  (m+n)~.
\ee
For a single quark corresponding to the fundamental representation
($m=1$, $n=0$) of QCD, we get $C_2 = (N_c^2 - 1)/(2 N_c) = 4/3$. 

To compute the cubic Casimir we require the set of $l_i = n_i+(N-1)/2$,
\bea
l_1 &=& \frac{2m+n}{3} + 2 \\
l_2 &=& \frac{-m+n}{3} +1 \\
l_3 &=& \frac{-m-2n}{3} \\
\sum_i l_i &=& N \\
\sum_i l_i^2 &=& N + \sum_i n_i^2 = 
  \frac{2}{3} \left( m^2 +n^2+mn\right) + 2 (m+n) +5 \\
\sum\limits_{i<j} l_i l_j &=& l_1(l_2+l_3) + l_2 l_3 = 
  2 - \frac{2m+n}{3} - \frac{4m^2+n^2+4mn}{9} - \frac{3m+6n-nm+2n^2-m^2}{9} \\
&=&
2 - \frac{m^2+n^2+mn+3m+3n}{3} \\
\sum_i l_i^3 &=& \frac{2m^3-2n^3+3m^2n-3mn^2}{9}+3m^2+n^2+2mn+7m+5n+9~.
\eea
These can then be used to construct the cubic Casimir
following~\cite{Dubath:2002qv} and lead to
\be
C_3 (m,n) \equiv  {1\over 18} \left( m + 2 n +3\right) \left(n + 2 m
+3\right)\left(m-n\right)~.
\ee
Constructing $C_4 (m,n)$ requires the vectors $n_i^4$ and can be done
similarly to $C_2 (m,n)$; this leads to 
\bea
C_4 (m,n) &\equiv & {1\over 9} \left(m^4 + n^4 + 2 m n^3 + 2 m^3
n + 3 m^2n^2\right) + {2\over 3}
\left(m^3 + n^3 + 2 m^2 n + 2 m n^2\right) \nonumber \\
&&+ {1\over 6}\left(5 m^2 + 5 n^2 + 11 m n\right) 
- {1\over 2} \left(m + n\right)~.
\eea

%---------------------------------------------------------------------

\end{document}